
\topmatter
\title
On the self-similar solutions of normality equation
in two-dimensional case.
\endtitle
\rightheadtext{On the self-similar solutions of normality
equation}
\author
Boldin A.Yu.
\endauthor
\address
Bashkir State University,
Department of mathematics,
32, Frunze street, Ufa,
Russia, 450074
\endaddress
\email
yavdat\@bgua.bashkiria.su
\endemail
\date
December 10, 1993
\enddate
\abstract
     The integrability in quadratures of normality equation for
spatially homogeneous dynamical systems in two-dimensional space
is shown. The classical symmetries of this equation are
calculated and the corresponding self-similar solutions are
found.
\endabstract
\endtopmatter
\document
\head
1. Introduction.
\endhead
     The concept of dynamical systems accepting the normal shift
was introduced in \cite{1} and \cite{2}. This concept separates
the special subclass of Newtonian dynamical systems of the form
$$
\ddot\bold{r} = \bold{F}(\bold{r},\dot\bold{r})
\tag{1.1}
$$
preserving the orthogonality of trajectories and hypersurfaces
shifted along these trajectories. In \cite{1 - 8} such dynamical
systems were investigated in consecutively complicating
geometrical situations: first in $\Bbb R^2$ and $\Bbb R^n$ then
on Riemannian and Finslerian manifolds. \par
     The dynamical system \thetag{1.1} is called spatially
homogeneous
if its force field $\bold{F}$ does not depend on the point
$\bold{r} \in \Bbb{R}^n$ and if it contains only the dependence
on the
velocity vector $\bold{v} =\dot{\bold{r}}$.
In two-dimensional case let us denote the modulus of the velocity
vector by $v$ and the angle between the velocity vector and some
fixed direction in space by $\theta$. Let $\bold{N} =
\vert\bold{v}\vert^{-1} \bold{v}$ be a unit vector directed along
the velocity vector and let $\bold{M}$ be a unit vector such that
$\bold{M}\perp\bold{N}$. So in the terms of $\bold{N}$ and
$\bold{M}$ we have the following expansion for the force field
$\bold{F}$ of
spatially homogeneous system
$$
\bold{F}(v,\theta)=A(v,\theta)\bold{N}(\theta) +
B(v,\theta)\bold{M}(\theta)\tag{1.2}
$$
i.e. the force field \thetag{1.2} is uniquely defined by
coefficients $A$ and $B$ of this expansion. In \cite{1} and
\cite{2} it was shown that the coefficients $A$ and $B$ for the
force field $\bold{F}$ of the dynamical system accepting the
normal shift in $\Bbb{R}^2$ the following equations
$$
\gathered
B = -A_\theta\\
A A_\theta - v A A_{\theta v} + A_\theta A_{\theta \theta} =
-v A_\theta A_v
\endgathered
\tag{1.3}
$$
should be fulfilled. The second equation from \thetag{1.3} is the
simplest equation from numerous normality equations obtained for
the different geometrical situations in \cite{1 - 8}. It is the
main object of study in this paper. \par
\head
2. The integrability in quadratures.
\endhead
     To prove the integrability in quadratures of normality
equation \thetag{1.3} let's make the following substitutions in
it
$$
A = e^a \qquad v = e^w
\tag{2.1}
$$
As the result of substitution \thetag{2.1} we get the following
equation
$$
a_\theta - a_{\theta w} - a_\theta a_w +
a_\theta a_{\theta\theta} + a_\theta^3 = -a_\theta a_w\tag{2.2}
$$
The equation \thetag{2.2} assumes the reduction of order. Indeed,
setting $a_\theta = b$ into \thetag{2.2} we obtain the following
quasilinear equation of first order respective to the function
$b(w,\theta)$
$$
b b_\theta - b_w + b + b^3 = 0
\tag{2.3}
$$
According to the general theory of quasilinear differential
equations (see \cite{9}) the solution of equation \thetag{2.3} is
an implicit function defined from the equation
$$
\varPhi(C_1,C_2) = 0
\tag{2.4}
$$
where $C_1$ and $C_2$ are first integrals of characteristic
equation
$$
C_1 = \theta + \arctan(b) \qquad C_2 = \frac{b^2 e^{-2w}}{1+b^2}
\tag{2.5}
$$
and $\varPhi = \varPhi(x,y)$ is an arbitrary sufficiently smooth
function. The integrability in quadratures is proved. \par
\head
3. Classical symmetries and self-similar solutions.
\endhead
     In spite of the fact that the equation \thetag{1.3} is
integrable in quadratures the formula \thetag{2.4} does not lead
to sufficiently effective explicit formulae for the solutions of
equation \thetag{1.3} in general case. Therefore the problem of
investigation of symmetry group of equation \thetag{1.3} and
constructing the self-similar solutions of this equation remains
actual one. \par
     One-parameter group of classical symmetries  of the equation
$$
A A_\theta - v A A_{\theta v} + A_\theta A_{\theta\theta} =
-v A_\theta A_v
\tag{3.1}
$$
from \thetag{1.3} is one-parameter family of transformations of
three-dimensional space
$\sigma_\varepsilon:(v,\theta,A)\longrightarrow(\tilde v,\tilde
\theta,\tilde A)$ which is a group with respect to the
composition $\sigma_\varepsilon\circ\sigma_\delta =
\sigma_{\varepsilon+\delta}$. This group  conserves the equation
\thetag{3.1} unchanged. It is well-known that one-parameter
groups of transformation are related to the vector fields. The
following theorem takes place for the classical symmetry groups
of the equation \thetag{3.1}.
\proclaim{Theorem 1} Any one-parameter group of classical
symmetries of the equation \thetag{3.1} is defined by the
following vector field
$$
X = \alpha v \frac{\partial {}}{\partial v} + \beta
\frac{\partial {}}{\partial \theta} +
f(\ln\,v) A \frac{\partial {}}{\partial A}
\tag{3.3}
$$
where $\alpha$ and $\beta$ are some arbitrary constants and
$f = f(w)$ is some arbitrary function of one variable.
\endproclaim
     The proof of this theorem is based on the standard methods
(see \cite{10} and \cite{11}). \par
     Let's make the following substitution $v=e^w$ in the equation
\thetag{3.1} then we obtain the equation
$$
A A_\theta - A A_{\theta w} + A_\theta A_{\theta\theta} =
-A_\theta A_w\tag{3.3}
$$
In the new independent variables $w$ and $\theta$ vector field
from \thetag{3.2} get the following form
$$
X = \alpha\frac{\partial {}}{\partial w} + \beta\frac{\partial {}}
{\partial \theta} + f(w) A \frac{\partial {}}{\partial A}
\tag{3.4}
$$
Let $F(w)$ be an antiderivative of $f(w)$ from (3.4). Then
one-parameter group of classical symmetries of equation (3.3)
corresponding to the vector field \thetag{3.4} with $\alpha\ne0$
is the transformation of the form
$\sigma_\varepsilon:(w,\theta,A)\longrightarrow(\tilde w,\tilde
\theta,\tilde A)$ where
$$
\align
\split
&\tilde w = \tilde{w}(w,\theta,A,\varepsilon) = w + \alpha\,
\varepsilon\\
&\tilde\theta = \tilde{\theta}(w,\theta,A,\varepsilon) = \theta +
\beta\,\varepsilon\\
&\tilde A = \tilde{A}(w,\theta,A,\varepsilon) = A\,\exp(Z)
\endsplit
\tag{3.5}\\
&Z(w) = \frac{F(w + \alpha \varepsilon) - F(w)}{\alpha}
\tag{3.6}
\endalign
$$
In that case when $\alpha=0$ this transformation has another form
$$
\aligned
&\tilde w = \tilde{w}(w,\theta,A,\varepsilon) = w\\
&\tilde \theta = \tilde{\theta}(w,\theta,A,\varepsilon) = \theta +
\beta\varepsilon\\
&\tilde A = \tilde{A}(w,\theta,A,\varepsilon)
\endaligned
\tag{3.7}
$$
\definition{Definition 1}. The solution $A=\psi(w,\theta)$ of
equation \thetag{3.3} is called self-similar solution or
invariant solution with respect to the group \thetag{3.5} if the
following relation
$$
\tilde {A}(w,\theta,\psi) =
\psi(\tilde(w,\theta,\psi),\tilde{\theta}(w,\theta,\psi))
\tag{3.8}
$$
is fulfilled identically for each value of variables $w$ and
$\theta$.
\enddefinition
     First let's consider the case $\alpha\ne0$. Let
$A=\psi(w,\theta)$ be the self-similar solution of \thetag{3.3}.
Introduce the following notation
$$
\varphi(w,\theta) = \psi(w,\theta)\,\exp\left(-\frac{F(w)}
{\alpha}\right)
\tag{3.9}
$$
Substituting \thetag{3.5} into \thetag{3.8} and taking into
account \thetag{3.6} we obtain the following formula for the
function $\varphi(w,\theta)$ defined by \thetag{3.9}
$$
\varphi(w,\theta) = \varphi(w+\alpha\,\varepsilon,\theta+\beta\,
\varepsilon)\tag{3.10}
$$
which holds for all values of $w,\theta$ and $\varepsilon$. \par
     Substitute $\varepsilon = - w/\alpha$ into the equation
\thetag{3.10}. As result of this substitution we can get
$$
\varphi(w,\theta) = \varphi(0,\theta-\beta/\alpha\,w)
\tag{3.11}
$$
According to \thetag{3.11} function $\varphi(w,\theta)$ of two
variables is expressed via function of one variable. As it's
known such function is called the function of traveling wave
type. Now returning back to the function $\psi(w,\theta)$
according to \thetag{3.9} we obtain the explicit form of
self-similar solution of the equation \thetag{3.3} (if
$\alpha\ne0$)
$$
A = \psi(w,\theta) = \varphi(\theta-\beta/\alpha\,w)\,
\exp\left(\frac{F(w)}{\alpha}\right)\tag{3.12}
$$
Function $b(w,\theta)$ from \thetag{2.3} is the logarithmical
derivative of $\psi(w,\theta)$ from (3.12) with respect to
$\theta$. For it we have
$$
b(w,\theta) = \xi(\theta - \beta/\alpha\,w)
\tag{3.13}
$$
Function $b(w,\theta)$ from \thetag{3.13} is a solution of
traveling wave type of equation \thetag{2.3}. Moreover
$\xi(\theta)=\partial_\theta {\ln\,(\varphi(\theta))}$, note that
function $b(w,\theta)$ is a solution of functional equation
\thetag{2.4}. It's convenient to suppose that equation
\thetag{2.4} is resolved with respect to $C_1$
$$
C_1 = h(C_2)
\tag{3.14}
$$
Taking into account \thetag{2.5} substitute \thetag{3.13} into
\thetag{3.14} and denote by $p=\theta-\beta/\alpha\,w$ it's own
primary argument for the function $\xi(p)$. As result we obtain
the following equation for this function
$$
p+\arctan(\xi(p))=h(q\,e^{-2w})-\beta/\alpha\,w
\tag{3.15}
$$
where $q$ is a quantity independent on $w$
$$
q=\frac{\xi^2(p)}{1+\xi^2(p)}
$$
Differentiating \thetag{3.15} with respect to $w$ for the fixed
$p$ and then setting $w = 0$ we get an equation for the function
$h(q)$
$$
\frac{\partial h}{\partial q}=-\frac{\beta}{2 \alpha q}
$$
which is easy to be integrated. Its solution has the form
$$
h(q)=-\frac{\beta}{2 \alpha}\ln\,q+C_1
$$
where $C_1$ is an integration constant. Substituting this
expression into \thetag{3.14} then for $\xi(p)$ we obtain the
equation
$$
p+\arctan(\xi(p)) = -\frac{\beta}{2 \alpha}
\ln\left(\frac{\xi^2(p)}{1+\xi^2(p)}\right)+C_1
\tag{3.16}
$$
For the case when $\alpha\ne0$ we have proved the following
theorem. \proclaim{Theorem 2} Each self-similar solution of
normality equation \thetag{3.3} which is invariant with respect
to one-parameter group of classical symmetries in the case of
$\alpha\ne0$ has the form
$$
A(\theta,w)=C_2 \exp\left(\int_0^{\theta-(\beta/\alpha)w} \xi(p)\,dp
\right)\,\exp\left(\frac{F(w)}{\alpha}\right)
\tag{3.17}
$$
where function $\xi(p)$ is defined from \thetag{3.16} and $F(w)$
is an antiderivative of $f(w)$ from \thetag{3.4}.
\endproclaim
Note that $C_1$ and $C_2$ in \thetag{3.16} and \thetag{3.17} are
the integration constants different from first integrals from
\thetag{2.4} and \thetag{2.5}. \par
     Let's consider the following case: $\alpha=0$ and
$\beta\ne0$ in \thetag{3.4}. In this case one-parameter group of
classical symmetries is defined by the formula \thetag{3.7}. The
condition of self-similarity for the solution $A=\psi(w,\theta)$
of the equation \thetag{3.3} has the form
$$
\psi(w,\theta)\,\exp(f(w)\,\varepsilon)=\psi(w,\theta+
\beta\,\varepsilon)\tag{3.18}
$$
It's identically fulfilled for any $w,\theta$ and $\varepsilon$.
So assuming that $\varepsilon = -\theta/\beta$ and substituting
such $\varepsilon$ into \thetag{3.18} we find
$$
\psi(w,\theta)=\varphi(w)\,\exp\left(\frac{f(w)\,\theta}{\beta}
\right)\tag{3.19}
$$
where $\varphi(w)=\psi(w,0)$. Function $b(w,\theta)$ in
\thetag{2.3} is a logarithmical derivative of the function
$\psi(w,\theta)$ with respect to $\theta$. For this function we
have the following representation
$$
b(w,\theta)=\frac{f(w)}{\beta}
\tag{3.20}
$$
Function $b(w,\theta)$ from \thetag{3.20} does not depend on
$\theta$; it is the solution of functional equation \thetag{2.4}
which is convenient to take resolved with respect to the second
argument $C_2$
$$
C_2=h(C_1)
\tag{3.21}
$$
Taking into account formula \thetag{3.20} and the explicit form
of the first integrals \thetag{2.5} we rewrite \thetag{3.21} as
follows
$$
\frac{f^2(w)\,e^{-2w}}{\beta^2+f^2(w)}=
h\left(\theta+\arctan\left(\frac{f(w)}{\beta}\right)\right)\tag{3.22}
$$
Left hand side of \thetag{3.22} does not depend on $\theta$ hence
right hand side also should not depend on $\theta$. It means that
function $h=h(\theta)$ is an identical constant. From
\thetag{3.22} for $f(w)$ one can obtain
$$
f(w)=\pm\,\sqrt{\frac{\beta h\,e^{2w}}{1-h\,e^{2w}}}
\tag{3.23}
$$
\proclaim{Theorem 3} In the case of $\alpha=0$ normality equation
\thetag{3.3} has self-similar solution of the form \thetag{3.19}
with an arbitrary function $\varphi(w)$ invariant under the
action of one-parameter group of classical symmetries
\thetag{3.7} if and only if functional parameter $f(w)$ in
\thetag{3.7} has the form \thetag{3.23}.
\endproclaim
     Case $\alpha=\beta=0$ and $f(w)\ne0$ completes the study of
self-similar solution of normality equation \thetag{3.3}. This
case is a trivial one. The equation \thetag{3.8} for the group of
classical symmetries \thetag{3.7} has an identically zero
self-similar solution only.\par
     Author thanks B\.I\.Suleimanov who found the substitution
\thetag{2.1}. He is grateful to R\.A\.Sharipov for setting up the
problem and useful discussions. Author thanks Russian Fund of
Fundamental Researches for the financial support (project
93-011-273). \par

\enddocument